\documentclass[aps,prl,twocolumn,superscriptaddress,showpacs]{revtex4-1}
\usepackage{amsmath,amssymb,wasysym,graphicx}
\usepackage[utf8]{inputenc}
\usepackage[T1]{fontenc}
\usepackage{xcolor}

\IfFileExists{newtxtext.sty}
{\usepackage{newtxtext,newtxmath}}
{\IfFileExists{stix.sty}
	{\usepackage{stix}}
	{\IfFileExists{mathptmx.sty}
		{\usepackage{mathptmx}}{} } }

\usepackage{textcomp}

\usepackage{bm}

\IfFileExists{siunitx.sty}{\usepackage{booktabs,siunitx}}{}

\usepackage{color}
\definecolor{LinkColor}{rgb}{0.256,0.439,0.588}
\usepackage{hyperref}
\hypersetup{
	pdfauthor={good guys},
	pdftitle={good title},
	colorlinks=true,
	citecolor=LinkColor,
	linkcolor=LinkColor,
	urlcolor=LinkColor
}

\newcommand{\beq} {\begin{equation}}
\newcommand{\eeq} {\end{equation}}
\newcommand{\bea} {\begin{eqnarray}}
\newcommand{\eea} {\end{eqnarray}}
\newcommand{\be} {\begin{equation}}
\newcommand{\ee} {\end{equation}}

\newcommand{\ket}[1]{\left|#1\right>}
\newcommand{\bra}[1]{\left<#1\right|}

\def\Eq#1{Eq.~(\ref{#1})}
\def\Fig#1{Fig.~\ref{#1}}
\def\avg#1{\left\langle#1\right\rangle}

\begin{document}
\title{Non-Hermitian strongly interacting Dirac fermions: a quantum Monte-Carlo study}

\author{Xue-Jia Yu}
\affiliation{International Center for Quantum Materials, School of Physics, Peking University, Beijing 100871, China}

\author{Zhiming Pan}
\affiliation{Institute for Theoretical Sciences, WestLake University, Hangzhou 310024, China}

 \author{Limei Xu}
\affiliation{International Center for Quantum Materials, School of Physics, Peking University, Beijing 100871, China}
\affiliation{Collaborative Innovation Center of Quantum Matter, Beijing, China}
\affiliation{Interdisciplinary Institute of Light-Element Quantum Materials and Research Center for Light-Element Advanced Materials, Peking University, Beijing, China}

\author{Zi-Xiang Li}
 \email{zixiangli@iphy.ac.cn}
 \affiliation{Beijing National Laboratory for Condensed Matter Physics,
	and Institute of Physics, Chinese Academy of Sciences, Beijing 100190, China}
\affiliation{School of Physical Sciences, University of Chinese Academy of Sciences, Beijing 100049, China}
\date{\today}

\begin{abstract}
 Exotic quantum phases and phase transition in the strongly interacting Dirac systems has attracted tremendous interests. On the other hand, non-Hermitian physics, usually associated with dissipation arising from the coupling to environment, emerges as a frontier of modern physics in recent years. In this letter, we investigate the interplay between non-Hermitian physics and strong correlation in Dirac-fermion systems. We develop a sign-problem-free projector quantum Monte-Carlo (QMC) algorithm for the non-Hermitian interacting fermionic systems. Employing state-of-the-art projector QMC simulation, we decipher the ground-state phase diagram of the Honeycomb Hubbard model in the presence non-Hermitian asymmetric spin resolved hopping processes. Intriguingly, the antiferromagnetic ordering induced by Hubbard interaction is enhanced by the non-Hermitian asymmetric hopping. More remarkably, our study reveals that critical properties of the quantum phase transition between Dirac semi-metal and AF ordered phases are consistent with the XY universality class in Hermitian system, implying Hermiticity is emergent at the quantum critical point. The numerically-exact QMC approach utilized in this study is easily applied to other non-Hermitian interacting fermionic models, hence paving a new avenue to investigating quantum many-body physics in non-Hermitian systems.  
 
\end{abstract}

\maketitle

%\section{INTRODUCTION}%
%\label{sec:introduction}
\emph{Introduction.}---Fathoming various exotic quantum phases and phase transition triggered by strong correlation between electrons is one of the central issues in modern condensed matter physics~\cite{sachdev2011quantum}. In particular, inspired by the experimental realization of graphene\cite{GrapheneReview} and topological phases\cite{Qi2011RMP, Kane2010RMP}, the interaction driving spontaneously symmetry breaking(SSB) phases and the associated quantum phase transition in Dirac fermions attracts growing interests. Since most strongly correlated systems are theoretically intractable in more than one-dimension, numerical approaches play vital roles in understanding Dirac-fermion systems in the presence of strong electronic interaction. Extensive numerical studies on interacting Dirac systems reveal an abundance of intriguing phenomena arising from the interplay between Dirac physics and strong electronic interaction, including interaction driven topological and other exotic phases of matter\cite{Lang2013prl, Wu2011PRB, Assaad2011PRL, Raghu2008PRL, Honerkamp2015PRB, Vishwanath2012PRL, Herbut2013PRB, Franz2010PRB, Li2017PRB}, Gross-Neveu quantum criticality\cite{sorella2012absence,Li2015NJP, Wang2014NJP,  Jian2017PRB, Herbut2006PRL, Tang2018Science, Assaad2017PRL, Herbut2016PRB, Assaad2015PRB, Sarma2014PRB, Herbut2013PRX, Lu2012PRB,Abolhassan2022PRL,Lang2019PRL,Li2018SA,yu2022prb} and continuous phase transition beyond Landau's paradigm~\cite{Li2017NC,li2019deconfined,li2020prb,liao2022prb_b}.

In recent years, understanding non-Hermitian (NH) physics in quantum systems emerges as a frontier invoking considerable attention~\cite{ashida2020non,bender2007making,lee2014prx}. Non-Hermitian physics arises if the system is coupled to the environment in the presence of dissipation or measurement~\cite{daley2014quantum,Jean1992PRL,El-Ganainy2018,photonicsReview}. Additionally, non-Hermitian band Hamiltonian provides a conceptually effective description of the quasi-particle with finite life-time resulting from electron-electron/electron-phonon interaction or disorder scattering\cite{Fu2018PRL,Fu2020PRL}. Previous studies demonstrate that plentiful exotic phenomena arise from non-Hermiticity, for instance skin effect\cite{yao2018prl,Fang2020PRL,song2019prl,kunst2018prl,jiang2019prb,xiao2020non,Yang2020PRL,Fang2022NC,ZhaiPRR2022,Wang2022arXiv}, exotic topological phases~\cite{bergholtz2021rmp,xiao2017observation,zhou2018observation,zeuner2015prl,ghatak2018prb,zhou2019prb,kawabata2019prx,kawabata2019topological,gong2018prx,kawabata2018prb,yao2018bprl,shen2018prltopological,xue2020prl,kawabata2019prx,zhou2019prb,Ghatak_2019,lee2019prl,okuma2020prl,Chen2022PRB,ZhaiPRB2018} and novel quantum critical behaviours~\cite{ashida2016pra,ashida2017parity,kawabata2017prlinformation,yamamoto2021universal,hayata2021dynamical,pakrouski2021prr,hamazaki2019prlmbl,alsallom2021fate,pan2020non,jian2021prl,jian2021prb,he_arxiv2209,ortega_arxiv2211,hsieh_arxiv2211,chang2020prr}, which have not been established in Hermitian systems. Most early studies on non-Hermitian physics focus on single-particle systems, whereas very recently, increasing attentions have been paid in the quantum many-body effect in the non-Hermitian systems~\cite{yamamoto2019prl,nakagawa2018prl,louren2018prbkondo,yoshida2018prb,orito_arxiv2212,Garcia_arxiv2211,wang_arxiv2210,ifmmode2022prr,suthar2022prb,chen_arxiv2202,kawabata2022prb,garca2022prx,orito2022prb,zhang_arxiv2109,tsubota2022prb,hyart2022prr,sun2022biorthogonal,zhai2020prb,shackleton2020prr,takasu_arxiv2004,matsumoto2020prl,tu_arxiv2203,tu2022scipost,mak_arxiv2301,chen_arxiv2208,wang2022prb,tzeng2021prr,kattel_arxiv2301,Han_arxiv2302}. 

\begin{figure}[tb]
\includegraphics[width=0.48\textwidth]{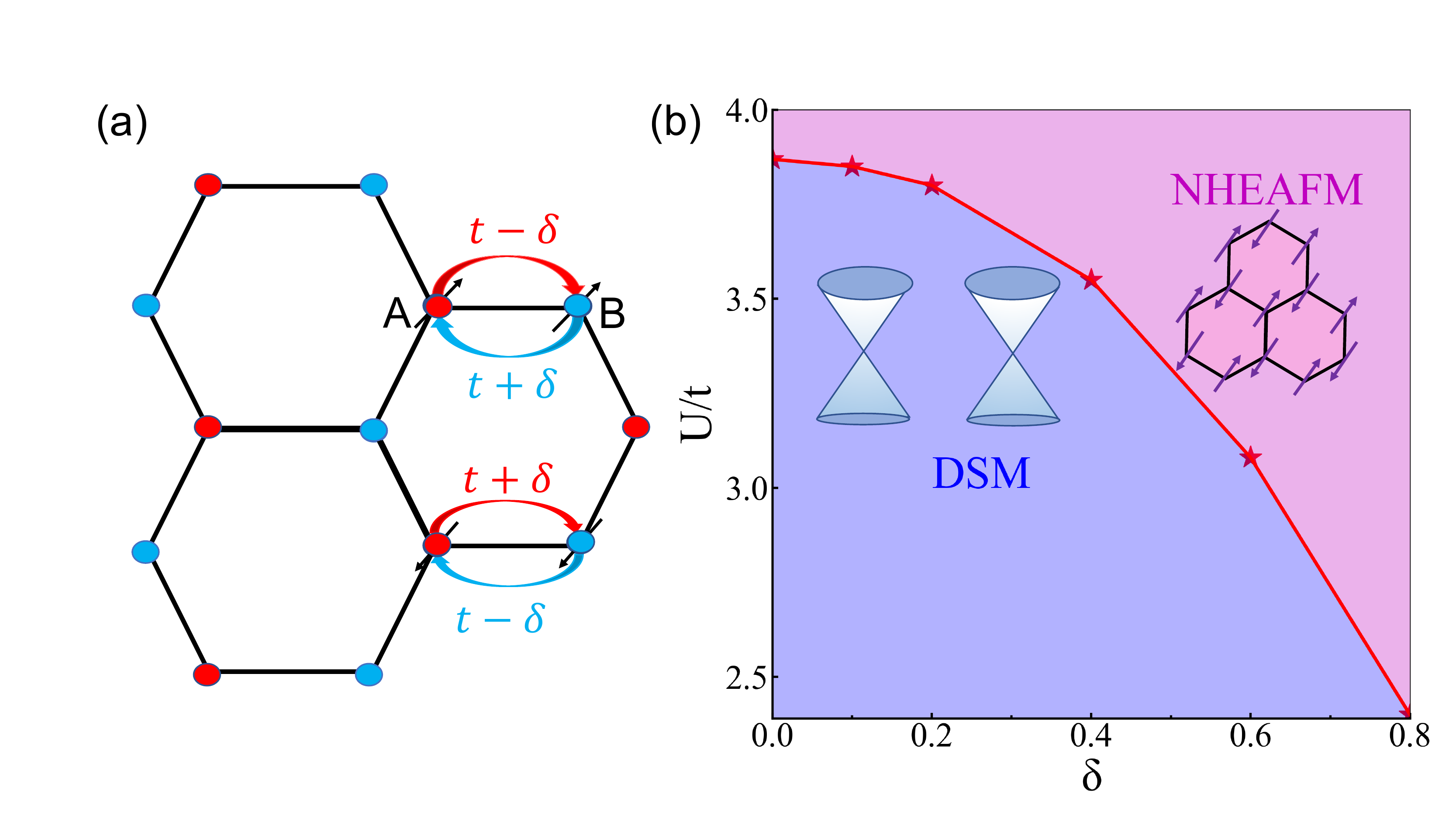}
\caption{Schematic asymmetry hopping (a) and ground state phase diagram (b) of non-Hermitian interacting model on the honeycomb lattice. In (a), the red(blue) circles represent the sites in A(B) sublattice. In (b), NHEAFM denotes non-Hermitian enhanced antiferromagnetism ordered phase and DSM denotes Dirac semi-metal phase. The red line is the phase boundary between DSM and NHEAFM phase, and red star points are numerical results obtained from projector QMC simulations.}
\label{fig:phase_diagram}
\end{figure}

Nevertheless, the effects of interplay between strong correlation and Dirac fermions in non-Hermitian systems remain unexplored hitherto. Moreover, the intrinsically unbiased numerical algorithm applying to the non-Hermitian quantum many-body models in more than one-dimension is lacking. To address these crucial issues, we systematically study the ground-state properties of a non-Hermitian model featuring Dirac fermions in the presence of Hubbard interaction. For the first time, we develop a non-Hermitian version of projector QMC, applying to investigating the ground-state properties of non-Hermitian interacting fermionic systems. Remarkably, in the framework of the developed algorithm, we construct a non-Hermitian Hubbard model free from the notorious sign problem by virtue of time reversal symmetry\cite{wu2005prb, Li2017PRL,li2015prb,Wang2015PRL,Xiang2016PRL,li2019sign}, enabling the large-scale QMC simulation without numerical approximation. Employing the numerically exact QMC simulation, we access the ground-state phase diagram of the model. The main results of the state-of-the-art QMC simulation are summarized as follows: 1. The interaction induced AFM order in Honeycomb Hubbard model is robust in the presence of non-Hermiticity. Intriguingly, the AFM order is enhanced by the asymmetric non-Hermitian process. 2. The numerical results reveal a continuous quantum phase transition occurring from DSM to AFM ordered phase. Surprisingly, despite the presence of non-Hermitician terms in the Hamiltonian, the quantum phase transition between Dirac semimetal (DSM) and antiferromagnetic (AFM) phases belongs to the Hermitian version of chiral-XY universality class\cite{ostmeyer2020prb,Li2017NC,otsuka2018prbxy}. Hence, the QMC results reveal the emergence of Hermiticity at the quantum phase transition point. Additionally, we present a systematic renormalization-group (RG) field theory analysis of the effective low-energy theory describing the non-Hermitian DSM-AFM transition, providing an understanding of the numerical results unveiled by QMC simulation. Notably, the QMC approach developed for non-Hermitian system is a general algorithm to tackle the non-Hermitian interacting fermionic models. We believe that our study paves a new route to investigating the quantum many-body physics in non-Hermitian systems.

\begin{figure}[tb]
\includegraphics[width=0.5\textwidth]{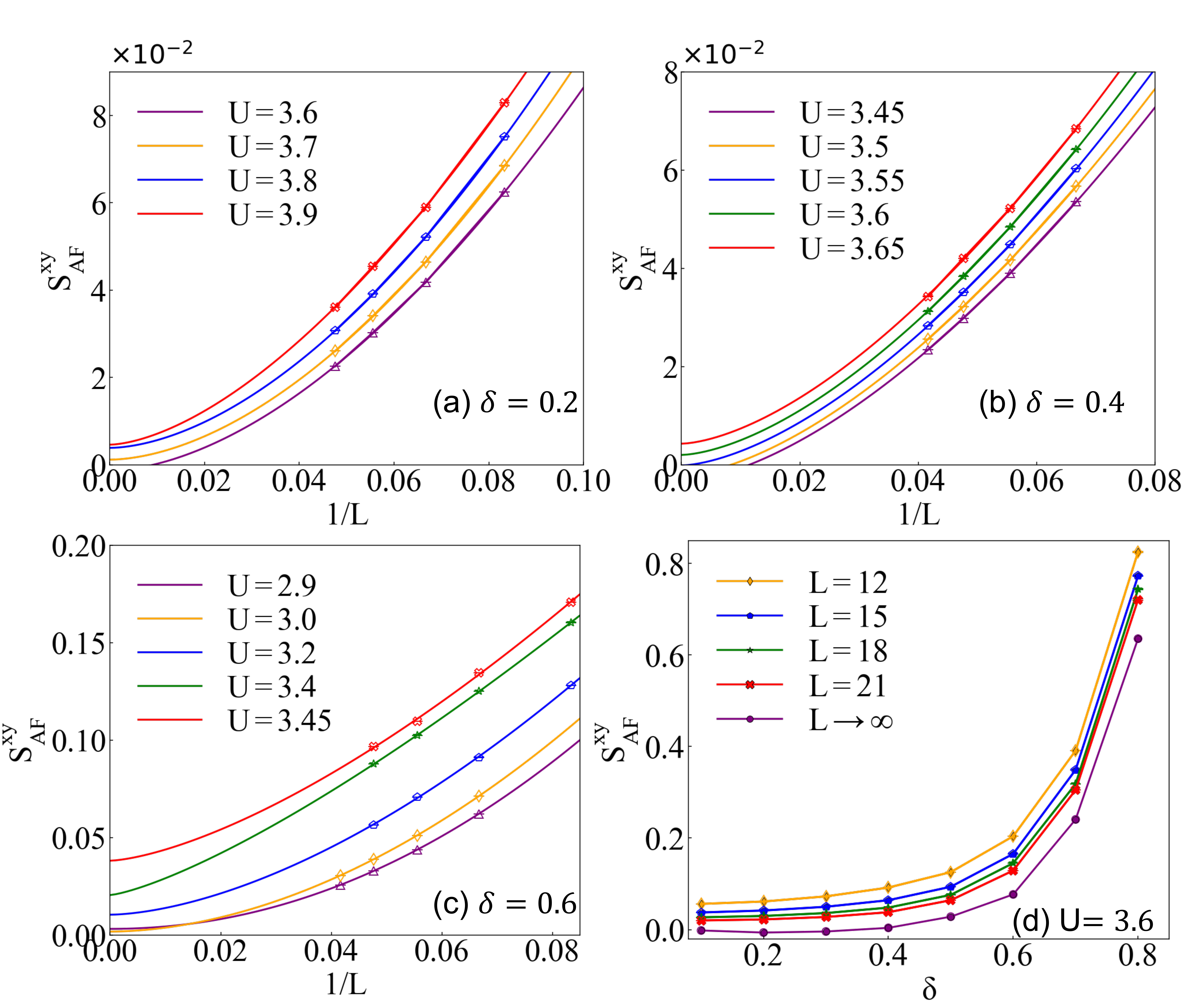}
\caption{Finite-size scaling of XY-AFM static structure factor $S^{XY}_{AF}$ at different non-Hermitian parameter $\delta$ and  Hubbard interaction $U$ close to the DSM-AFM transition.  (a) $\delta = 0.2$, $U_{c} \approx 3.8$. (b) $\delta = 0.4$, $U_{c} \lesssim 3.6$. (c) $\delta = 0.6$, $U_{c} \lesssim 3.2$. The largest system size in the simulation is $L=24$. The more accurate results of $U_{c}$ are accessed from RG-invariant ratio, as shown in \Fig{fig4}. The results of $U_c$ obtained by different approaches are approximately consistent with each other. (d) XY-AFM structure factor $S_{AF}^{XY}$ as a function of $\delta$ for $U/t=3.6$. The results for finite size $L=12,15,18,21$, and extrapolated to thermodynamic limit $L \rightarrow \infty$ by the polynomial fitting of  $S_{AF}^{XY}$ versus $1/L$ as $S_{AF}^{XY} = a + b/L^{c}$ for $U/t = 3.6$.}
\label{fig:fss_afm}
\end{figure}

%\section{Model and methods}%
%\label{sec:model}
\emph{Model and methods.}---In this paper, we study the non-Hermitian extension of an interacting model on the honeycomb lattice\cite{sachdev2011quantum}. Specifically, as shown in Fig.\ref{fig:phase_diagram} (a), we introduce the asymmetric hopping to construct a non-Hermitian version of the Hubbard model\cite{hayata2021prb}:
\begin{equation}
\begin{split}
\label{E1}
&H = H_0 + H_1 = \\
& -t \sum_{\langle i ,j \rangle, \sigma} c^{\dagger}_{i,\sigma}c_{j,\sigma}+h.c + \sum_{i}U(c^{\dagger}_{i\uparrow}c_{i\uparrow}-\frac{1}{2})(c^{\dagger}_{i\downarrow}c_{i\downarrow}-\frac{1}{2}) \\
& -\delta \sum_{\langle i, j \rangle} (c^{\dagger}_{i,\uparrow}c_{j,\uparrow}-c^{\dagger}_{j,\uparrow}c_{i,\uparrow}-c^{\dagger}_{i,\downarrow}c_{j,\downarrow}+c^{\dagger}_{j,\downarrow}c_{i,\downarrow}),
\end{split}
\end{equation}
where $H_1$ and $H_0$ are the non-Hermitian and Hermitian Hamiltonian (Hubbard model),
respectively. $\langle i j \rangle$ refers to the nearest-neighboring (NN) sites $i$ and $j$. $c^{\dagger}_{i,\sigma}$ creates a fermion on-site $i$ with spin $\sigma=\uparrow,\downarrow$.
$t,\delta$ are symmetric and asymmetric hopping amplitude, respectively. For $\delta=0$, Eq \ref{E1} is reduced to Hermitian Hubbard model on the honeycomb lattice, and has been extensively studied\cite{sorella2012absence,Assaad2015PRB}. We set our energy units $t = 1$ throughout this article. $U$ is conventional on-site Hubbard interaction. The Hermiticity of the model is broken when asymmetric hopping is introduced. The model is $\mathcal{PT}$ symmetric\cite{Bender1998prl}, rendering the possibility that the Hamiltonian exhibits real eigenvalues despite its non-Hermiticity. For Hamiltonian \Eq{E1}, in the non-interacting limit $U=0$ the eigenstates preserve $\mathcal{PT}$-symmetry if $\delta$ is smaller than the exceptional point (EP) $\delta_c=1$, at which eigenstates of the Hamiltonian coalesce. In the presence of Hubbard interaction, the exact diagonalization calculations in small clusters suggest the EP $\delta_c=1$ is stable under interaction, namely when $\delta<\delta_c$ the eigenvalues of the interaction Hamiltonian \Eq{E1} are real (see Sec.IV in Supplementary Materials (SM) for the details).

To investigate the ground-state properties of the non-Hermitian model, we generalize the algorithm of PQMC. To our knowledge, it is the first approximation-free quantum Monte-Carlo algorithm to study the ground-state properties of quantum many-body non-Hermitian model.  Intriguingly, even though in the presence of non-Hermiticity,  the model at half filling is free from the notorious sign problem at half filling in the framework of our algorithm, with the details introduced in SM I. Thus, the model offers a promising platform to investigate the interplay between non-Hermiticity and quantum many-body effects. In this work, we perform state-of-the-art QMC simulations\cite{bss1981prd,ss1981prb,li2019sign} to study the ground-state phase diagram of the model. We focus on the model at half filling, and explore the quantum phase and the critical properties of the interaction driven quantum phase transition in non-Hermitian system.

%\section{Quantum phase diagram}%
%\label{sec:phase}
\emph{Quantum phase diagram.}---Before presenting the details of PQMC results, we briefly summarize the most salient features of the ground-state phase diagram of the model \ref{E1}. The schematic phase diagram with varying asymmetric hopping amplitude $\delta$ and on-site Hubbard interaction $U$ is shown in Fig.\ref{fig:phase_diagram}(b). For $\delta = 0$, a quantum phase transition from DSM to AFM phase is driven by the Hubbard interaction. We obtain critical point $U_{c} = 3.87$, which is consistent with previous studies in the literature\cite{sorella2012absence}. Our large-scale QMC simulations convincingly show that AFM order is robust, and more appealingly, strongly enhanced in the presence of non-Hermiticity. The transition point $U_c$ from DSM to the non-hermitian enhanced antiferromagnetic (NHEAFM) phase is shifted to a smaller value with increasing asymmetric hopping amplitude. The enhancement of AFM ordering by non-Hermitian asymmetric hopping is beyond our conventional understanding that non-Hermiticity is expected to destroy long-range ordering, and constitute a main discovery of our numerical simulation. Furthermore, we employ standard finite-size scaling (FSS) procedure to extract the critical exponents and determine universality class of the DSM-NHEAFM transition. Even though the lattice model is non-Hermitian, the results reveal that the transition between DSM and NHEAFM belongs to \emph{Hermitian} chiral XY universality class \cite{ostmeyer2020prb,Li2017NC,otsuka2018prbxy}, implying Hermiticity is emergent at the DSM-NHEAFM transition point at low-energy limit.

\emph{Non-Hermitian enhanced antiferromagnetism}---To investigate the effect of non-Hermiticity on AFM long-range ordering triggered by Hubbard interaction, we compute structure factor of AFM order, the definition of which is introduced in the section I of SM. Notice that non-Hermitian asymmetric hopping reduces the spin rotational $SU(2)$ symmetry down to a $U(1)$ corresponding to spin rotation in the $x-y$ plane, leading to the consequence that AFM structure factor in $z$-direction $S_{AF}^{z}$ is not equivalent to that in $xy$-plane $S_{AF}^{xy}$. The numerical results unequivocally point out, AFM order in $xy$-plane is dominant over the ordering in $z$-direction in the presence of non-Hermitian asymmetric hopping, as shown in the section IV of SM. Hence, in the following we present the numerical results of AFM ordering in $xy$-plane.   

For the Hermitian limit $\delta=0$, the standard FSS procedure yields the critical point of DSM-AFM transition $U_c = 3.87$, in agreement with extensive studies in previous literature. In the presence of non-Hermitian asymmetric hopping, the FSS analysis also establishes existence of AFM long-range ordering triggered by strong Hubbard interaction. \Fig{fig:fss_afm}(a)-(c) depicts the AFM structure factors $S_{AF}^{xy}$ for various interaction parameters and linear system sizes $L$, at several non-Hermitian asymmetric hopping parameters $\delta=0.2,0.4,0.6$, clearly indicating quantum phase transition occurs from a DSM to NHEAFM ordered phase with increasing Hubbard interaction strength. More remarkably, the critical value of Hubbard $U$ for the DSM-NHEAFM transition decreases upon increasing non-Hermitian hopping parameter $\delta$, suggesting the enhancement of AFM ordering by non-Hermiticity.   

%\begin{figure}[tb]
%\includegraphics[width=0.5\textwidth]{Fig3.pdf}
%\caption{XY-AFM structure factor $S_{AF}^{XY}$ as a function of $\delta$ for $U/t=3.6$. The results for finite size $L=12,15,18,21$, and extrapolated to thermodynamic limit $L \rightarrow \infty$ by the polynomial fitting of  $S_{AF}^{XY}$ versus $1/L$ as $S_{AF}^{XY} = a + b/L^{c}$ for $U/t = 3.6$.}
%\label{fig3}
%\end{figure}

To explicitly observe the enhancement of AFM long-range order at thermodynamic limit, we present the 
 AFM structure factor as a function of $\delta$ for different linear system sizes $L$ and the extrapolated results to $L \rightarrow \infty$ ( \Fig{fig:fss_afm}(d)). The Hubbard interaction parameter is fixed at $U=3.6$. \Fig{fig:fss_afm}(d) clearly indicates AFM structure factors monotonically increases with $\delta$. For $U=3.6$, the ground state is DSM with non-Hermitian hopping $\delta=0$, which is verified by the extrapolated result of AFM structure factor $S_{AF}^{xy}(L \rightarrow \infty) = 0$. With increasing $\delta$, the AFM ordering is enhanced, resulting in a quantum phase transition from DSM to AFM ordred phase driven by non-Hermitian hopping, occurring at $\delta_c \approx 0.4$. Consequently, the numerical results unambiguously demonstrate the enhancement of AFM ordering by non-Hermiticity in model \Eq{E1}.

\begin{figure}[tb]
\includegraphics[width=0.5\textwidth]{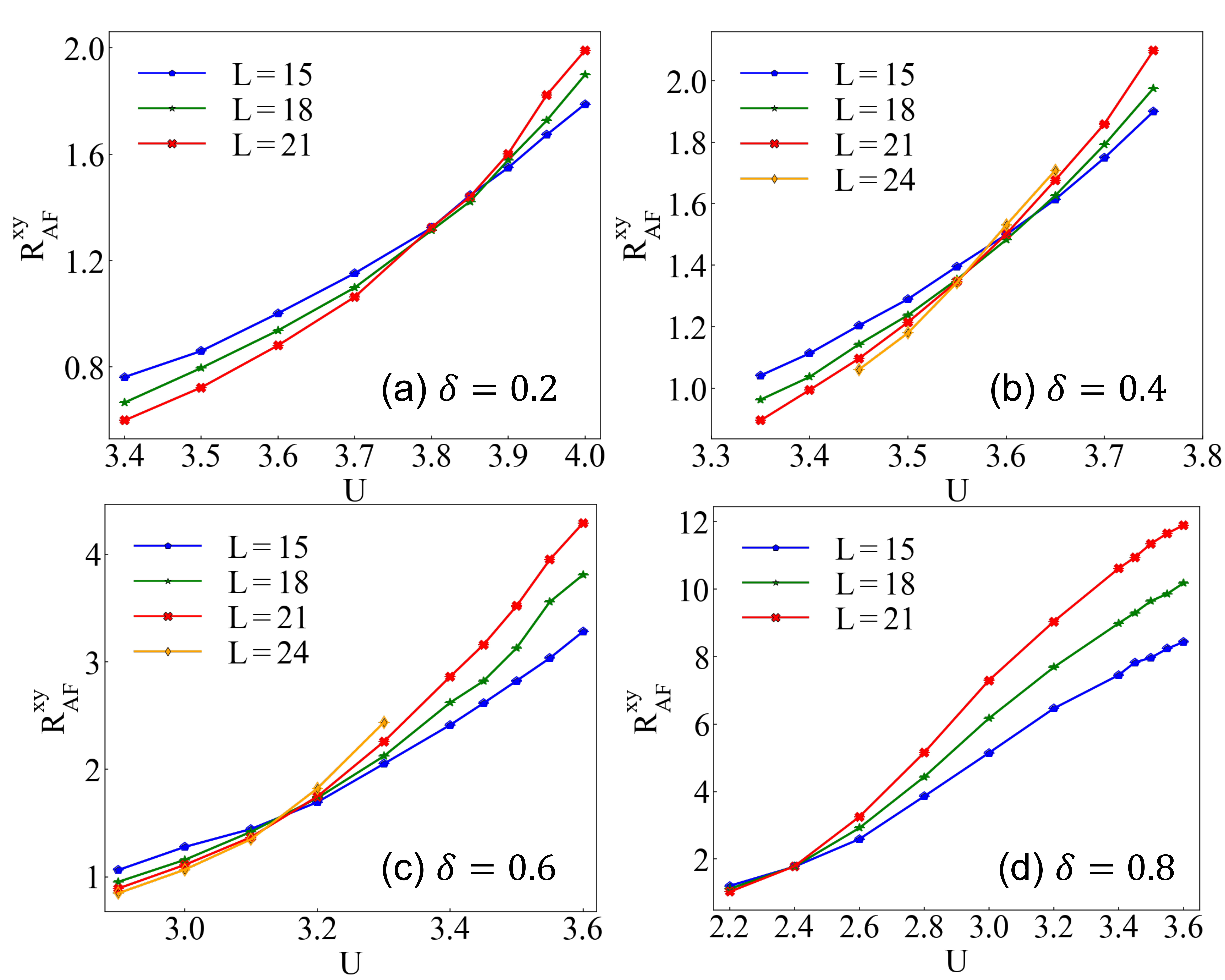}
\caption{RG-invariant ratio for XY-AFM order as a function of Hubbard interaction strength $U$ for different $\delta$. (a) $\delta=0.2$, $U_{c} \approx 3.80 $. (b) $\delta=0.4$, $U_{c} \approx 3.55$. (c) $\delta=0.6$, $U_{c} \approx 3.08 $. (d) $\delta=0.8$, $U_{c} \approx 2.4 $. }
\label{fig4}
\end{figure}

\begin{figure*}[tb]
\includegraphics[width=1.0\textwidth]{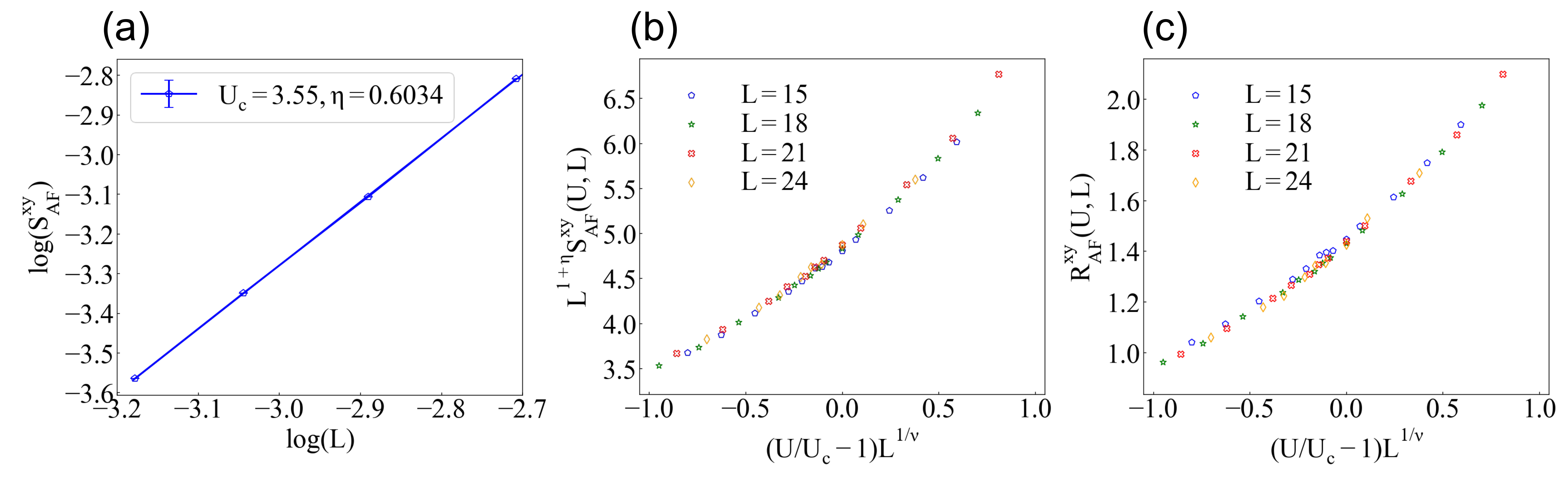}
\caption{Finite-size scaling analysis for the QCP between DSM and NEAFM phases.  (a) Log-log plot for XY-AFM structure factors versus system size at QCP. The slope of the fitted line yields anormalous dimension $\eta=0.603\pm0.03$.
(b) The data collapse analysis for the rescaled XY-AFM structure as a function of $(U/U_{c}-1)L^{1/\nu}$ at $\delta=0.4$ for system sizes $L=15,18,21,24$.  
(c) The data collapse analysis for the RG-invariant ratio of XY-AFM structure factors as a function of $(U/U_{c}-1)L^{1/\nu}$ at $\delta=0.4$ for system sizes $L=15,18,21,24$. The results of  critical exponents $\eta = 0.603\pm0.03$ and $\nu=1.07\pm0.02$ are consistent with chiral-XY universality class within erro bar.}
\label{fig5}
\end{figure*}

To determine accurate transition point between DSM and AFM ordered phases, we compute the RG-invariant correlation-length ratio for AFM order defined as: 
\begin{equation}
\label{E2}
R(L)_{AF}^{XY} = \frac{S_{AF}^{XY}(\vec{Q},L)}{S_{AF}^{XY}(\vec{Q}-\delta\vec{q})}-1,
\end{equation}
where $\vec{Q}=(0,0)$ labels the momentum at which the structure factor is maximum for AFM order. $\delta\vec{q} = (\frac{2\pi}{L},\frac{2\pi}{L})$ is a minimal momentum on lattice shift from $\vec{Q}$. In the long-range ordered phase, correlation length ratio increases with system size, while if the AFM order is short-ranged, the trend is opposite. At the critical point, the  RG-invariant correlation-length ratio is independent on system size owing to the nature of scaling invariant, therefore offering a powerful theoretical approach to identify critical point of phase transition. In \Fig{fig4}, we present the results of correlation-length ratios as a function of Hubbard $U$ for several values of $\delta$, giving rise to the DSM-NHEAFM quantum critical points, which are approximately consistent with the results given by the finite-size scaling of AFM structure factors. Consequently, we obtain the ground-state phase diagram of Hubbard model on Honeycomb lattice in the presence of non-Hermitian asymmetric hopping, explicitly revealing the AFM ordering associated with electronic interaction is strongly enhanced by the non-Hermitian asymmetric hopping in \Eq{E1}.

%\section{EMGERNT CHIRAL XY UNIVERSALITY CLASS}%
%\label{sec:class}
\emph{Emergent chiral XY universality class.}---Having accessing the ground-state phase diagram of the non-Hermitian model \Eq{E1}, we investigate the critical property of the quantum phase transition between DSM and NHEAFM phases. In the absence of non-Hermitian hopping, DSM-AFM transition featured in Hubbard model on Honeycomb lattice is extensively investigated, belonging to chiral-Heisenberg universality class~\cite{Ladovrechis2023prb}. As aforementioned, the non-Hermitian hopping introduced in \Eq{E1} breaks the spin $SU(2)$ symmetry to $U(1)$, is thus expected to change the universality class of the phase transition. In Hermitian system, the continuous quantum phase transition between DSM and the in-plane AFM ordered phase breaking $U(1)$ symmetry is described by the Gross-Neveu-Yukawa model, and belongs to chiral-XY universality class, as revealed by field-theory analysis and numerical simulation on lattice models. In this section, we decipher the critical properties of the DSM-AFM transition in the presence non-Hermitian asymmetric hopping process.

In the regime sufficiently close to critical point $U_c$, the AFM structure factors obeys the scaling relation: 
\begin{equation}
\label{E4}
S_{AF}^{XY} = L^{-(1+\eta)}F_{1}[(U-U_{c})L^{1/\nu}],
\end{equation}
where $F_1$ is a unknown function, $\eta$ and $\nu$ are critical exponents determining the critical properties of the phase transition, where $\eta$ is anomalous dimension of AFM order parameter and $\nu$ is correlation length exponent. Other critical exponents are available to deduce from $\eta$ and $\nu$ by hyper-scaling relation.  To extract anomalous dimension $\eta$, we perform log-log plot of AFM structure factor at quantum critical point versus system size, and the anomalous dimension of AFM order parameter is given by the slope: $S_{AF}^{XY}(L,U_c) = a L^{-1-\eta}$. We present the result of extrapolation with fixed $\delta=0.4$ in \Fig{fig5}(a),  yielding the anomalous dimension $\eta= 0.603 \pm 0.03 $. Then we perform the standard data collapse procedure to obtain the critical exponent $\nu$. There exists an appropriate value of $\nu$ such that the points $((U-U_c)L^{1/\nu},S_{AF}^{XY} L^{1+\eta})$ at various $U$ should collapse into a single curve for different linear system sizes $L$. We present the results of data collapse for $\delta=0.4$ in \Fig{fig5} (b), which renders the correlation length exponent $\nu = 1.07\pm0.02$. The results of critical exponents $\eta = 0.603\pm0.03$ and $\nu = 1.07\pm0.02$ are consistent with the numerical results of Gross-Nevue transition belonging to Hermitian chiral-XY universality class on Hermitian interacting lattice model\cite{ostmeyer2020prb,Li2017NC,otsuka2018prbxy}. To further verify the numerical results of critical exponents, we perform the data collapse of RG-invariant correlation-length ratios satisfying the scaling relation: $R_{AF}^{XY} = F_{2}[(U-U_{c})L^{1/\nu}]$, using the result of $\nu = 1.07$. The results of $R_{AF}^{XY}$ for different system sizes exhibit excellent collapse, ensuring our analysis yields convincing results of critical exponents. Furthermore, with the choice of a different value of $\delta=0.6$, we implement the same procedure and extract critical exponents for the DSM-NHAFM transition. The results are included in the section V of SM, in agreement with the results for $\delta=0.4$.

To conclude, the non-Hermitian asymmetric hopping introduced in model \Eq{E1} breaks $SU(2)$ spin rotational symmetry to $U(1)$, favoring AFM ordering in $xy$-plane. Intriguingly, although in the presence of non-Hermiticity, DSM-NHEAFM transition realized in the model belongs to the same universality class as the Hermitian version of the phase transition, namely chiral-XY universality class.

%\section{LOW ENERGY EFFECTIVE THEORY}%
%\label{sec:fieldtheory}
\emph{Low-energy effective theory.}---We derive the effective Hamiltonian describing the low-energy Dirac fermions in the non-Hermitian model \Eq{E1} on honeycomb lattice (see Sec.VI in SM for the details). The result writes:
\begin{equation}
\label{S8}
h(\vec{q}) = -v_F(q_x s_{0}\otimes\sigma_{1}+q_y s_{0}\otimes\sigma_{2}- \frac{iq_y\delta }{t} s_{3}\otimes\sigma_{1}- \frac{iq_x\delta}{t} s_{3}\otimes\sigma_{2}),
\end{equation}
where $s,\sigma$ are Pauli matrices acting on spin and sublattice spaces respectively, and $v_F = \frac{3t}{2}$ is the Fermi velocity in the absence of non-Hermitian hopping, namely $\delta=0$. The direct diagonalization of the Hamiltonian yields the energy dispersion (see Sec.VI in SM), which is purely real and corresponds to a Dirac cone with renormalized Fermi velocity $\tilde{v}_{F} = v_{F}\sqrt{1-(\delta/t)^{2}}$. Since the density of state(DOS) near the energy of Dirac point is proportional to the inverse of Fermi velocity, the reduction of Fermi velocity results in the increase of DOS near the Dirac point, which offers a plausible explanation on the enhancement of AFM order as revealed by our numerical simulation. More crucially, to understand the emergence of Hermiticity at XY-AFM QCP as unambiguously revealed in our numerical results, we perform an one-loop perturbative renormalization group calculation on the Gross-Neveu low-energy effective theory (see the SM VII for details ). The beta function of the non-Hermitian parameter $\delta$ is $\frac{d\delta}{dl}=-2 \hat{g}^{2}\delta$ (see the SM VII for details), demonstrating the irrelevance of the non-Hermitian parameter at low-energy limit. Therefore, the Hermitian chiral XY universal class is emergent at the QCP. We leave a more systematic field-theory 
 study as a future work.

%\section{SUMMARY}%
%\label{sec:summary}
\emph{Conclusion and discussion.}---In summary, we propose an innovative QMC algorithm to investigate the ground-state properties of quantum many-body models in the presence of non-Hermiticity. We construct a non-Hermitian interacting fermionic model, more explicitly, Hubbard model with non-Hermitian asymmetric hopping. Remarkably, the model is free from the notorious sign problem, such that the ground-state properties of model with large system size are accessible. Employing state-of-the-art QMC simulation, we systematic investigate the quantum phases and phase transition emerging in this model. The results unambiguously reveal that the AFM ordering triggered by strong Hubbard interaction is enhanced by non-Hermitian asymmetric hopping. The interaction driven quantum phase transition from DSM to the NHEAFM phase belongs to the chiral-XY universality class as previously unveiled in Hermitian systems, although the non-Hermiticity is present in the model.   

In the perspective of experimental realization, the non-Hermitian hopping process is induced by the single-particle loss or gain on the bond associated with coupling to the environment, corresponding to Lindblad operator $L_{\avg{ij}} = c_i + c_j$ in the quantum master equation. When the quantum-jump terms are neglible, the resulting effective Hamiltonian involves the non-Hermitian hopping terms as discussed in our study. More remarkably, since the principles to guarantee sign-problem-free in QMC simulation proposed in Hermitian models are straightforwardly generalized to non-Hermitian algorithm, it is promising to design more non-Hermitian interacting fermionic models featuring fascinating physics based on our approach. For example, designing sign-problem-free models and investigating the interacting effect on the non-Hermitian topological phase is particularly intriguing. Hence, our study paves a new avenue towards understanding the interplay between quantum many-body and non-Hermitian physics in a theoretically controlled approach.

\begin{acknowledgements}
We thank Zijian Wang, Shang Liu for helpful discussions. We thank the computational resources provided by the TianHe-1A supercomputer, the High Performance Computing Platform of Peking University, China. X.-J.Y. and L.X. are supported by the National Natural Science Foundation of China under Grant No.11935002, and the National 973 project under Grant No. 2021YF1400501. Z.X.L acknowledges support from the
start-up grant of IOP-CAS.
  \end{acknowledgements}

\bibliographystyle{apsrev4-1}
\bibliography{NHDF_arxiv}

\clearpage
\onecolumngrid

\newpage
\begin{widetext}
			
\section{Supplemental Material for  Phase Diagram of Non-Hermitian Interacting Dirac Fermion}

\subsection{Section I: Projector Quantum Monte-Carlo algorithm for non-Hermitian fermionic model}
\label{sec:A2}

For QMC simulation on fermionic quantum many-body systems, projector QMC is a widely used algorithm to investigate the ground-state properties of the models. The conventional version of projector QMC algorithm is constraint for the Hermitian many-body models. We develop a generalized projector QMC algorithm to simulate the ground-state properties of the non-Hermitian interacting fermionic models. For non-Hermitian Hamiltonian, the eigenvalues are not guaranteed to be real. The observables associated with the ground state of a non-Hermitian Hamiltonian are evaluated as: 
\bea
\avg{O}_{G} = \frac{\langle \psi_{L,G}|\hat O|\psi_{R,G} \rangle}{\langle \psi_{L,G}|\psi_{R,G} \rangle}
\eea
In the scheme of projector QMC simulation, the ground state of an interacting Hamiltonian is achieved through projecting a trivial wave function. As long as the trivial wave function is not orthogonal to the ground-state wave function $\langle \psi_{T} | \psi_{G}\rangle \ne 0$, which is generically obeyed in quantum many-body systems, the ground state of a given Hamiltonian is accessed at sufficiently large projector parameter: $|\psi_{G} \rangle = \lim_{\Theta \rightarrow \infty} e^{-\Theta \hat H} | \psi_{T} \rangle$. Correspondingly, in the non-Hermitian Hamiltonian, the right and left ground states are accessed as: $|\psi_{R,G} \rangle = \lim_{\Theta \rightarrow \infty} e^{-\Theta \hat H} | \psi_{R,T} \rangle$ and $|\psi_{L,G} \rangle = \lim_{\Theta \rightarrow \infty} e^{-\Theta \hat{H}^\dagger} | \psi_{L,T} \rangle$, where $| \psi_{R,T} \rangle$ and $| \psi_{L,T} \rangle$ are right and left trial wave function. Thus, the expectation value of ground state for non-Hermitian Hamiltonian are evaluated as:
\bea
 \langle \hat O \rangle_{G} = \langle \psi_{L,G}|\hat O|\psi_{R,G} \rangle /\langle \psi_{L,G} | \psi_{R,G} \rangle=\lim_{\Theta \rightarrow \infty}\langle \psi_{L}|e^{-\Theta \hat H}\hat O e^{-\Theta \hat H}|\psi_{R}\rangle/\langle\psi_{L}|e^{-2\Theta\hat H}|\psi_{R}\rangle
 \label{ES2}
\eea
Consequently, we can implement the standard procedure of Trotter decomposition and Hubbard-Stratonovich transformation, as employed in the conventional projector QMC algorithm for Hermitian systems, to evaluate the ground-state expectation value of the observables in non-Hermitian systems in terms of \Eq{ES2}. Under Trotter decomposition, the projector parameter is decomposed into $N_\tau$ imaginary-time slices $\Delta_{\tau}$, where $\Delta_{\tau}=\Theta/N_\tau$. The Hubbard-Stratonovich transformation is utilized to decouple the four-fermion interacting terms into the bilinear-fermion operators coupled to auxiliary fields. 

We perform projector QMC simulations to investigate the ground-state properties of 
the non-Hermitian Hubbard model introduced in \Eq{E1} the main text. In our work, we choose right (left) trial wave function as the ground-state wave function of the non-interacting part of the model $H_{0}$ ($H_{0}^{\dagger}$).
 In our simulation, we fix projector parameter $\Theta=2L$, and have carefully checked the convergence of results against using larger values of $\Theta$. We employ the discrete H-S transformation to decouple the Hubbard interaction in non-Hermitian Hamiltonian \Eq{E1}:
 \bea
e^{-\frac{\Delta_\tau U}{2}(n_{i\uparrow}+n_{i\downarrow} ) ^2} = \sum_{l=\pm 1,\pm 2} \gamma(l) e^{i\sqrt{\frac{\Delta_\tau U}{2}}\eta(l)(n_{i\uparrow}+n_{i\downarrow})} 
\label{HS}
\eea
 with the four-valued parameters: $\gamma(\pm 1) = 1+\sqrt{6}/3$, $\gamma(\pm 2) = 1-\sqrt{6}/3$, $\eta(\pm 1) = \pm \sqrt{2(3-\sqrt{6})}$, $\eta(\pm 2) = \pm \sqrt{2(3+\sqrt{6})}$.
In projector QMC, we perform Trotter decomposition to discretize the imaginary time and the time step is $\delta \tau=0.1$ in our simulations. We have verified that $\Delta\tau$ is small enough to guarantee the convergence of results.

To characterize various long-range orderings, we evaluate structure factors for the corresponding order parameters. For model Eq.(1), the dominant orderings are AFM spin order breaking spin rotational symmetry. The structure factors for AFM orders with system size $N_s=2\times L \times L$ are defined as:

\begin{equation}
\begin{split}
\label{SF}
&S_{AF}^{XY} = \sum_{\alpha,\beta}[S_{AF}^{XY}]^{\alpha\beta} \\
&[S_{AF}^{XY}]^{\alpha \beta} = \frac{1}{L^{2}}\sum_{r,r^{\prime}}(-1)^{\alpha}(-1)^{\beta} \bra{\Psi_{0}} S^{+}_{r\alpha}S^{-}_{r^{\prime}\beta}+S^{-}_{r\alpha}S^{+}_{r^{\prime}\beta}\ket{\Psi_{0}}\\
&S^{zz}_{AF}=\sum_{\alpha,\beta}[S^{zz}_{AF}]^{\alpha \beta} \\
&[S^{zz}_{AF}]^{\alpha \beta} = \frac{1}{L^{2}}\sum_{r,r^{\prime}}(-1)^{\alpha}(-1)^{\beta} \bra{\Psi_{0}} S^{z}_{r \alpha}S^{z}_{r^{\prime} \beta}\ket{\Psi_{0}},
\end{split}
\end{equation}
where $r,r^{'}$ denote unit cells, $\ket{\Psi_{0}}$ is the ground state, $\alpha,\beta \in A,B$ are sublattice indices, $(-1)^{\alpha}=1(-1)$ for $\alpha$ = A(B), and we take the trace of the structure factor. 
%\subsection{APPENDIX A: SIGN PROBLEM FREE FOR $\mathcal{PT}$ SYMMETRIC NON-HERMITIAN HUBBARD MODEL}
\subsection{Section II: Sign problem free for $\mathcal{PT}$ symmetric non-Hermitian Hubbard model}
\label{sec:A1}
In this section, we show that the non-Hermitian Hamiltonian \Eq{E1} is $\mathcal{PT}$ symmetric, and more importantly, sign-problem-free in QMC simulation with the algorithm of generalized projector QMC for non-Hermitian systems. As shown in \Eq{E1} of main text, the model Hamiltonian reads: 
\begin{equation}
\begin{split}
\label{S1}
H = \sum_{\langle i,j \rangle}\{-(t+\delta)c^{\dagger}_{i\uparrow}c_{j\uparrow}-(t-\delta)c^{\dagger}_{j\uparrow}c_{i\uparrow}-(t-\delta)c^{\dagger}_{i\downarrow}c_{j\downarrow} 
-(t+\delta)c^{\dagger}_{j\downarrow}c_{i\downarrow}\}+\sum_{i}\frac{U}{2}(c^{\dagger}_{i\uparrow}c_{i\uparrow}-\frac{1}{2})(c^{\dagger}_{i\downarrow}c_{i\downarrow}-\frac{1}{2}) 
\end{split}
\end{equation}
Considering spatially inversion transformation under bond center $\mathcal{P}$ and time reversal operation  $\mathcal{T}: i K \sigma_y$, where $\sigma_y$ denotes second Pauli Matrix in spin space and $K$ is Hermitian conjugation. Obviously, the Hamiltonian \Eq{E1} is invariant under $\mathcal{PT}$ transformation. Thus, the model we considered is $\mathcal{PT}$ symmetric featuring many interesting properties discussed in the literatures~\cite{Bender1998prl,ruter2010observation}.

Under the discrete Hubbard-Stratonovich transformation defined in \Eq{HS}, the Hubbard interacting term is decoupled to the non-interacting operators coupled to the classical auxiliary fields depending on space and imaginary time. The decoupled Hamiltonian after H-S transformation at a given imaginary-time slice reads:
\bea
H_\tau = \sum_{\langle i,j \rangle}\{-(t+\delta)c^{\dagger}_{i\uparrow}c_{j\uparrow}-(t-\delta)c^{\dagger}_{j\uparrow}c_{i\uparrow}-(t-\delta)c^{\dagger}_{i\downarrow}c_{j\downarrow} 
-(t+\delta)c^{\dagger}_{j\downarrow}c_{i\downarrow}\}+
i\lambda \sum_i \sqrt{\frac{\Delta_\tau U}{2}}\eta(i,\tau) (n_{i\uparrow}+n_{i\downarrow})
\label{decoupled}
\eea
where $n_{i\uparrow(\downarrow)} = c^\dagger_{i\uparrow(\downarrow)}c_{i\uparrow(\downarrow)}$ is the electronic density operator on site $i$ with spin polarization in up(down)-direction. To prove the absence of sign problem, we introduce a partial particle-hole transformation as follows: 
\bea
c_{i\uparrow} \rightarrow d_{i\uparrow} \quad    c_{i\downarrow} \rightarrow d^\dagger_{i\downarrow} (-1)^{\alpha_i}
\eea
where $\alpha_i=\pm 1$ if the site $i$ belongs to A(B) sublattice. Under such transformation, the decoupled Hamiltonian \Eq{decoupled} reads:
\bea
H_\tau = \sum_{\langle i,j \rangle}\{-(t+\delta)d^{\dagger}_{i\uparrow}d_{j\uparrow}-(t-\delta)d^{\dagger}_{j\uparrow}d_{i\uparrow}-(t+\delta)d^{\dagger}_{i\downarrow}d_{j\downarrow} 
-(t-\delta)d^{\dagger}_{j\downarrow}d_{i\downarrow}\}+
i\lambda \sum_i \sqrt{\frac{\Delta_\tau U}{2}}\eta(i,\tau) (n_{i\uparrow}-n_{i\downarrow})
\eea
which is invariant under time-reversal symmetry $T=i\sigma_y K$. If the decoupled Hamiltonian under H-S transformation preserves an anti-unitary symmetry satisfying $T^2=-1$, the model is sign-problem-free in QMC simulation. Consequently, the non-Hermitian Hubbard model considered in our study is absent from sign problem in QMC simulation.

%\subsection{APPENDIX B: PROJECTOR QUANTUM MONTE CARLO}

\subsection{Section III: Exact diagonalization results}
It is easily shown that there exists an exact exceptional point at the $\delta=t$ in the model \ref{E1} in the absence of Hubbard interaction. In this section, we investigate the stability of exceptional points (EP) in the model \ref{E1} with Hubbard interaction. The Hamiltonian \ref{E1} preserves $\mathcal{PT}$ symmetry even in the presence of Hubabrd interaction. Nevertheless, the existence of interaction makes it infeasible to access the exceptional point straightforwardly. Hence, we perform exact diagonalization to calculate the eigenvalues of Hamiltonian \ref{E1}, and plot the distribution of eigenvalues in Fig.~\ref{SM5}. The horizontal and vertical axis represent real and imaginary parts of eigenvalues, respectively. As shown in Fig.~\ref{SM5}(${\rm{a-c}}$), when $\delta \le 1$, all the eigenvalues reside in the real axis, implying $\mathcal{PT}$ symmetry is not broken. Once the value of $\delta$ is increased to $\delta>1$, the complex eigenvalues of the Hamiltonian appear and $\mathcal{PT}$ symmetry is broken (see Fig.~\ref{SM5}.(${\rm{d-f}}$) for $\delta=1.05,1.1,1.2$). We have performed the calculation in various values of $U$ and the conclusion still holds. Consequently, our numerical results strongly suggest the exceptional point $\delta_c = 1$ is stable against Hubbard interaction in the model \ref{E1}. 

\begin{figure*}[tb]
\includegraphics[width=1.0\textwidth]{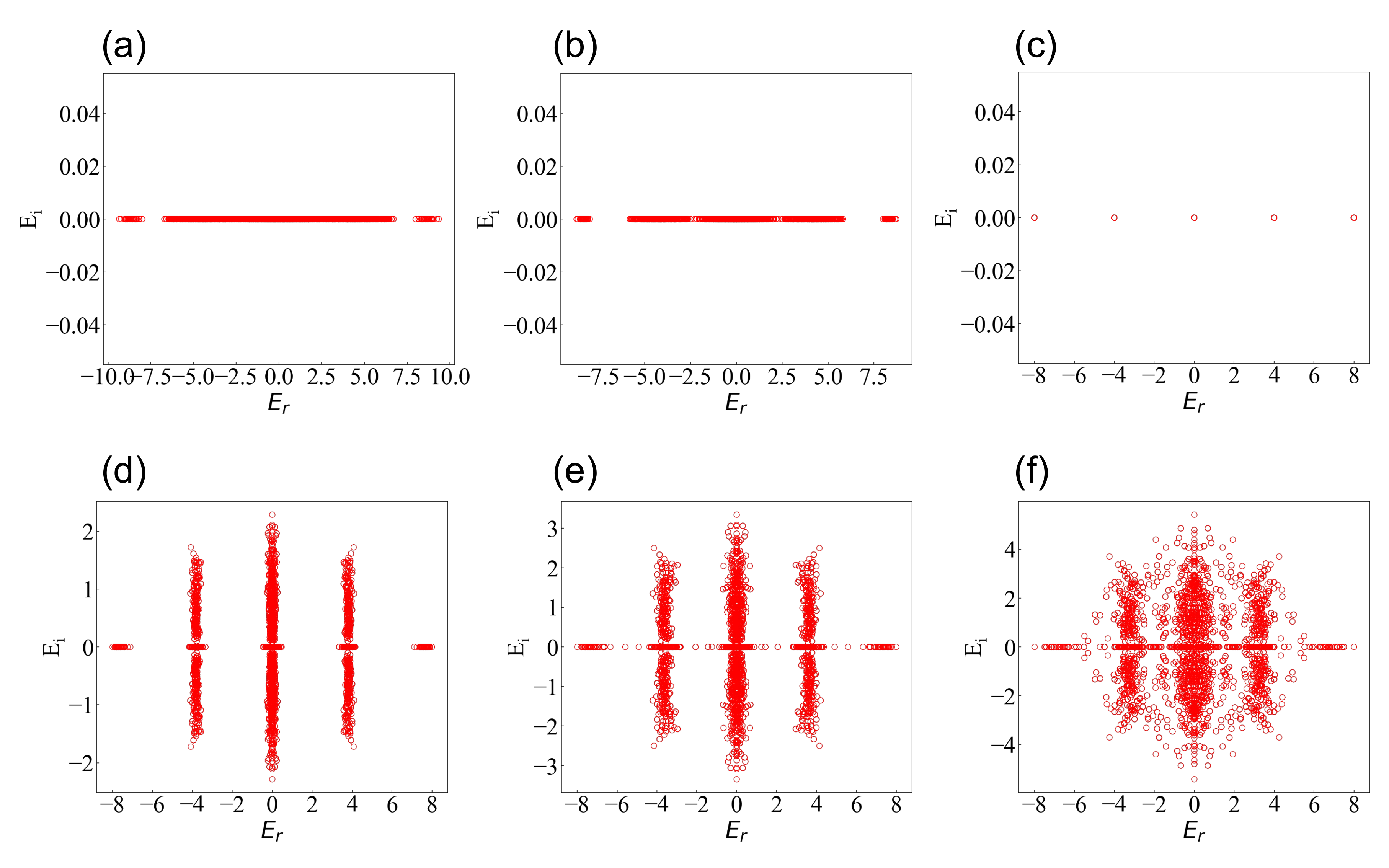}
\caption{ Exact diagonalization results for the different $\delta$. (a) $\delta=0.9$. (b) $\delta=0.95$. (c) $\delta=1.0$ (exceptional point). (d) $\delta=1.05$. (e) $\delta=1.1$. (f) $\delta=1.2$. All the energy spectrum are calculated with half-filled, the number of lattice sites in our simulation is $2\times2\times2$ on honeycomb lattice, and Hubbard interaction strength is set $U=4$. } 
\label{SM5}
\end{figure*}

\subsection{Section IV: QMC results of AFM structure factor in $z$-direction and $xy$-plane}
The non-Hermitian asymmetric hopping reduces the spin rotational $SU(2)$ symmetry down to a $U(1)$, corresponding to spin rotational symmetry in the $x-y$ plane, leading to the consequence that AFM structure factor in $z$ direction $S^{z}_{AF}$ is not equivalent to the one in $xy$-plane $S^{xy}_{AF}$. In this section, we explicitly show that AFM order in $xy$-plane is dominant over the ordering in $z$-direction owing to the existence of non-Hermitian asymmetric hopping. To this end, we calculate the transverse/longitudinal AFM structure factor as a function of Hubbard interaction strength $U$ for $\delta=0.4,0.6$, as shown in Fig~\ref{SM4}(a) and (b), respectively, unambiguously revealing that transverse AFM ordering is always stronger than the longitudinal one within the parameter regime under consideration, and the difference is more prominent with increasing non-Hermitian asymmetric hoppoing $\delta$. Hence, we conclude that in the presence of non-Hermitian asymmetric hopping, the $xy$-plane AFM ordering is dominant over AFM ordering in $z$-direction.

\begin{figure}[tb]
\includegraphics[width=0.8\textwidth]{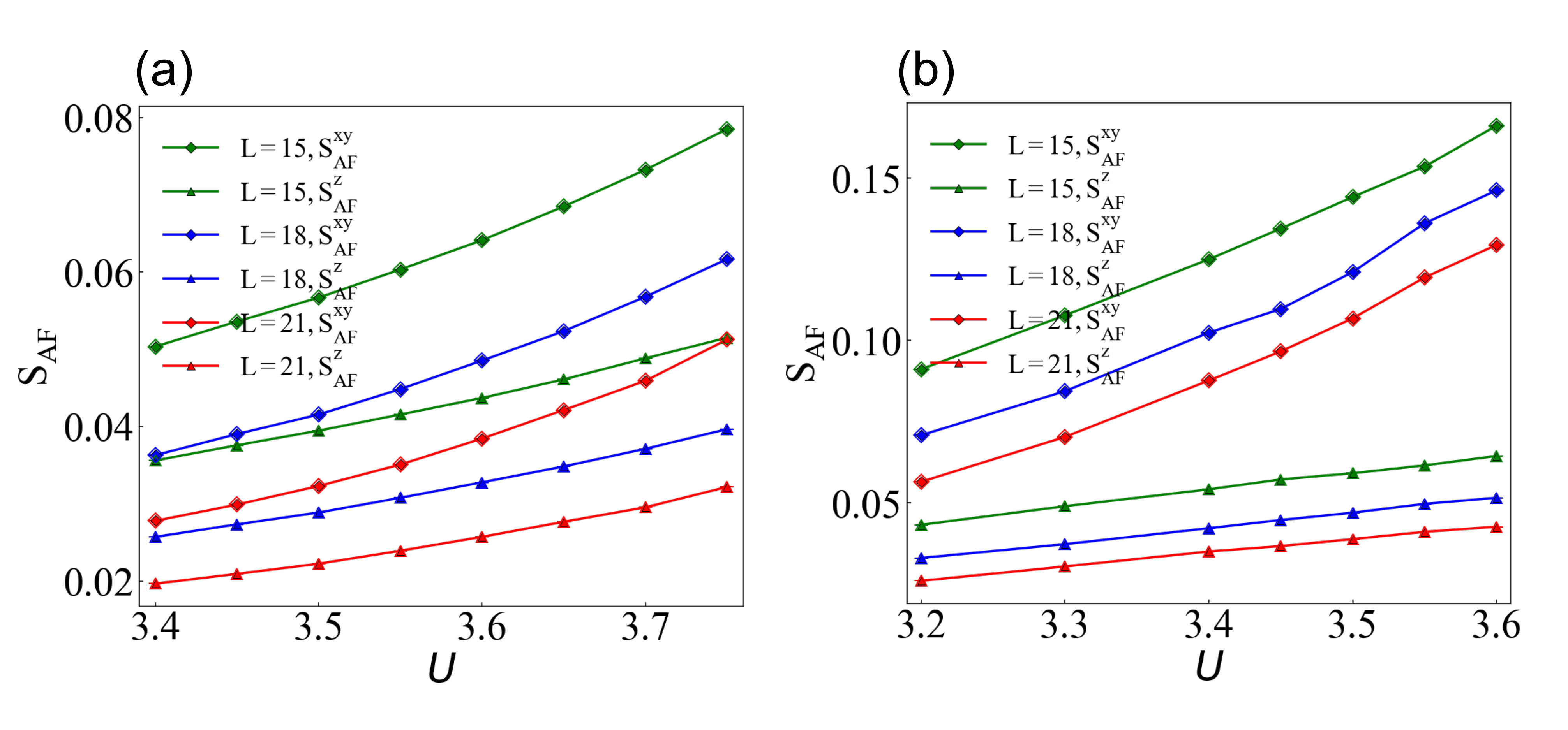}
\caption{Transverse/ longitudinal AFM structure factor $S_{AF}^{XY}$/$S_{AF}^{ZZ}$ as a function of Hubbard interaction $U$. (a) $\delta=0.4$, $L=15,18,21$. (b) $\delta=0.6$, $L=15,18,21$. We see the transverse AFM structure factor always stronger than longitudinal AFM structure factor within the interaction strength under consideration, which implies AFM order in $xy$-plane is dominant over the ordering in $z$-direction in the presence of non-Hermitian asymmetric hopping.}
\label{SM4}
\end{figure}

%\subsection{APPENDIX C: DATA COLLAPSE FOR $\delta = 0.6$}
\subsection{Section V: QMC results of critical properties for $\delta=0.6$}
\label{sec:A3}
In the maintext, we have investigated the critical properties of the quantum phase transition between Dirac semi-metal and AFM phase for $\delta=0.4$, and extract the critical exponents via the standard procedure of finite-size scaling analysis. Unexpectedly, the results reveal the critical exponents of the phase transition is consistent with the Hermitian chiral XY universality class within errorbars. In this section, we present the results of critical properties of DSM to NHAFM transition for $\delta=0.6$, confirming the conclusion of emergent Hermitian chiral XY universality class from the non-Hermitian lattice model.  

Fixing $\delta=0.6$, we present the results of data collapse analysis for AFM structure factor and AFM correlation-length ratio in \Fig{figS1}(a) and \Fig{figS1}(b), respectively. \Fig{figS1}(a) is the results of scaled AFM structure factors $L^{1+\eta}S_{AF}^{XY}$ as a function of $(U/U_{c}-1)L^{1/\nu}$ for different linear system sizes $L=15,18,21,24$. Given the critical exponents obtained from the analysis for the results of $\delta=0.4$, the scaled AFM structure factors for different system sizes collapse to a single smooth curve as a function of $(U/U_{c}-1)L^{1/\nu}$, indicating the DSM-NHAFM quantum phase transition for $\delta=0.6$ belongs to the same universality class with $\delta=0.4$, namely chiral-XY universality class. The same procedure of analysis is performed for correlation-length ratio. The results of correlation-length ratios for different linear system sizes obeys the scaling function with the choice of critical exponents. We also perform long-log plot of AFM structure factors versus linear system size $L$ at the quantum critical point $U_c = 3.08$, and the anomalous dimension $\eta = 0.6146$ is given by the linear fitting, which is consistent with $\eta =0.603 \pm0.03 $ for $\delta=0.4$ with error bar. Consequently, the phase transition from DSM to NHAFM phase for $\delta=0.6$ and $\delta=0.4$ belong to the same universality classes consisent with chiral XY transition, suggesting the emergent Hermiticity at the QCP between DSM and NHAFM phase in model \Eq{E1}.

\begin{figure}[tb]
\includegraphics[width=0.96\textwidth]{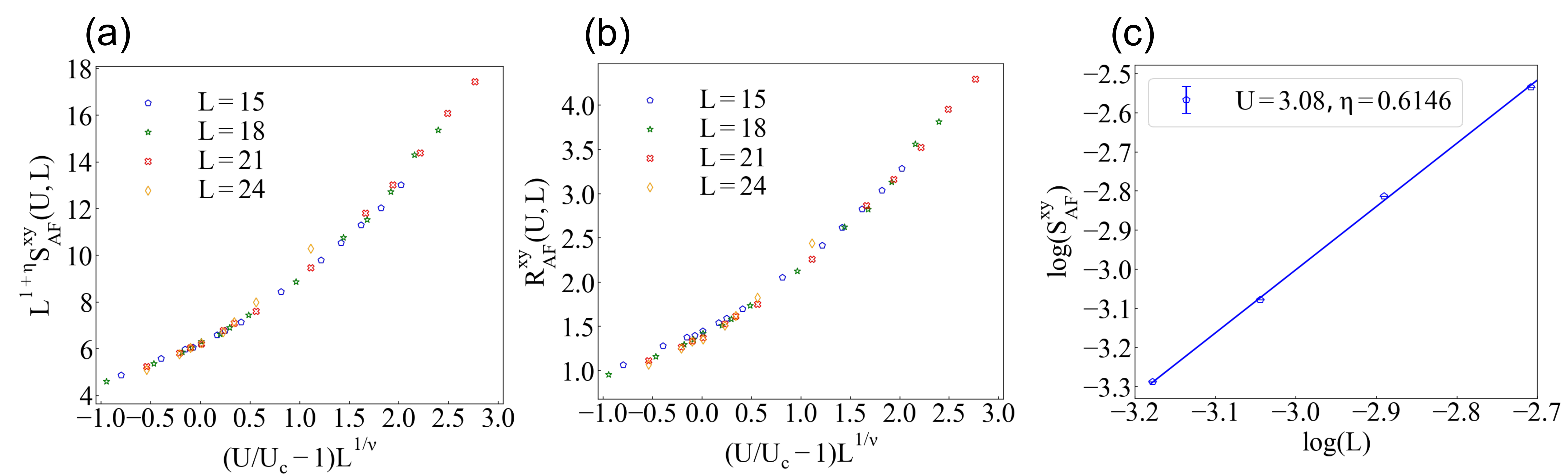}
\caption{(a) Transverse AFM structure factor $S_{AF}^{XY}$ defined in Eq.\ref{SF} as function of $(U/U_{c}-1)L^{1/\nu}$ at $\delta=0.6$ for different system sizes $L$. 
(b) Transverse AFM order RG invariant ratio $R_{AF}^{XY}$ defined in Eq.\ref{E2} as function of $(U/U_{c}-1)L^{1/\nu}$ at $\delta=0.6$ for different system sizes $L$. The QMC data are fully consistent with the critical point $U_{c} = 3.08$, exponents $\nu=1.073$, and $\eta = 0.6146$ of the chiral XY model
(c) Log-log plot for transverse AFM structure factor as a function of system size $L=15,18,21,24$ close to critical point.}
\label{figS1}
\end{figure}

%\subsection{APPENDIX D: DETAILS OF LOW ENERGY EFFECTIVE THEORY NEAR DIRAC PPINT}
\subsection{Section VI: Details of low energy effective theory near Dirac point}
\label{sec:A4}
In this section, we derive low-energy effective theory describing the quantum critical point separating DSM and NHAFM phases. First, we deduce the low-energy Hamiltonian of the model \Eq{E1} in the absence of Hubbard interaction. The non-interacting part of \Eq{E1} reads:
\begin{equation}
\begin{split}
\label{S5}
H =  -(t+\delta)\sum_{i,\alpha=1,2,3}a^{\dagger}_{i,\uparrow}b_{i+\kappa_\alpha,\uparrow}-(t-\delta)\sum_{i,\alpha=1,2,3}b^{\dagger}_{i+\kappa_\alpha,\uparrow}a_{i,\uparrow} -(t-\delta)\sum_{i,\alpha=1,2,3}a^{\dagger}_{i,\downarrow}b_{i+\kappa_\alpha,\downarrow}-(t+\delta)\sum_{i,\alpha=1,2,3}b^{\dagger}_{i+\kappa_\alpha,\downarrow}a_{i,\downarrow},
\end{split}
\end{equation}
where $a^{\dagger}_{i,\sigma},b^{\dagger}_{i,\sigma}$ with $\sigma=\uparrow,\downarrow$ creates electron with spin polarizaiton $\sigma$ at site $i$ on A and B sublattice, respectively.  $\kappa_1=\frac{a}{2}(1,\sqrt{3})$, $\kappa_2=\frac{a}{2}(1,-\sqrt{3})$, $\kappa_3=-a(1,0)$ ($a$ is lattice spacing) denote the three vectors of nearest-neighbor bonds on honeycomb lattice. Hereafter we set $a=1$ as unite length. 
After Fourier transform 
$a^{\dagger}_{i,\sigma}=\frac{1}{\sqrt{N/2}}\sum_{k}e^{i\vec{k}\vec{r}_{i}}a^{\dagger}_{k,\sigma}$, 
$b^{\dagger}_{i,\sigma}=\frac{1}{\sqrt{N/2}}\sum_{k}e^{i\vec{k}\vec{r}_{i}}b^{\dagger}_{k,\sigma}$
, where $N$ is number of sites and $\vec{k}$ belong to half of the first Brillouin zone, we arrive at the non-interacting Hamiltonian in momentum space $H=\sum_{k}\Psi^{\dagger}_{k}h(k)\Psi_{k}$, where we introduce four-component spinor  $\Psi^{\dagger}_{k}=(a^{\dagger}_{k,\uparrow},b^{\dagger}_{k,\uparrow},a^{\dagger}_{k,\downarrow},b^{\dagger}_{k,\downarrow})$. 
The matrix $h(\vec{k})$ is written as:
\begin{equation}
\label{S7}
h(\vec{k})=
\begin{pmatrix}
0 & -(t+\delta)\Delta^{*}(\vec{k}) & 0 & 0\\
-(t-\delta)\Delta(\vec{k}) & 0 & 0 & 0 \\
0 & 0 & 0 & -(t-\delta)\Delta^{*}(\vec{k})\\
0 & 0 & -(t+\delta)\Delta(\vec{k}) & 0
\end{pmatrix}
\end{equation}
and $\Delta(\vec{k})=\sum_{\alpha} e^{i\vec{k}\cdot\vec{\kappa}_\alpha}$. Expanding Hamiltonian near the Dirac point $\vec{K}=\frac{2\pi}{3\sqrt{3}}(\sqrt{3},1)$ and $\vec{k}=\vec{K}+\vec{q}$ where $|q| << 1$, we derive the low-energy effective Hamiltonian written as: 
\begin{equation}
\label{S8}
h_{\vec{K}}(\vec{q}) = v_F(-q_x s_{0}\otimes\sigma_{1}-q_y s_{0}\otimes\sigma_{2}+ \frac{iq_y\delta }{t} s_{3}\otimes\sigma_{1}- \frac{iq_x\delta}{t} s_{3}\otimes\sigma_{2}),
\end{equation}
where $s,\sigma$ are Pauli matrices acting on spin and sublattice spaces respectively, and $v_F = \frac{3t}{2}$ is the original Fermi velocity in the absence of non-Hermitian hopping, namely $\delta=0$. Diagonalization of \Eq{S8} gives rise to the energy dispersion of the Hamiltonian $E(\vec{q}) = \pm \frac{3}{2}\sqrt{t^{2}-\delta^{2}} \left|\vec{q}\right|$, corresponding to a Dirac cone with Fermi velocity $\tilde{v}_{F}=\frac{3\sqrt{t^{2}-\delta^{2}}}{2}$.  Hence,  the low-energy physics of the non-Hermitian Hamiltonian \Eq{E1} near the Dirac point $\vec{K}$ is described by a Dirac cone with renormalized Fermi velocity $\tilde{v}_{F}=\frac{3\sqrt{t^{2}-\delta^{2}}}{2}$. Similarly, employing the same procedure, we obtain the low-energy Hamiltonian near the Dirac point $\vec{K^{\prime}}=-\frac{2\pi}{3\sqrt{3}}(\sqrt{3},1)$:
\begin{equation}
\label{S9}
h_{\vec{K^{\prime}}}(\vec{q}) = v_F(q_x s_{0}\otimes\sigma_{1}-q_y s_{0}\otimes\sigma_{2}+ \frac{iq_y\delta }{t} s_{3}\otimes\sigma_{1}+ \frac{iq_x\delta}{t} s_{3}\otimes\sigma_{2}),
\end{equation}
the energy dispersion of which also describes a Dirac cone with Fermi velocity $\tilde{v}_{F}=\frac{3\sqrt{t^{2}-\delta^{2}}}{2}$. To conclude,    the effect of non-Hermitian hopping introduced in \Eq{E1} is reducing the Fermi velocity of the Dirac cone located at $\pm \vec{K}$ and thus leading to the consequence that density of state at low energy is increased, which offers a plausible explanation on the numerical results that AFM ordering is enhanced by the non-Hermitian hopping introduced in \Eq{E1}. 

\subsection{Section VII: Details of renormalization group calculations for the non-Hermitian Gross-Neveu-Yukawa model}
\label{sec:A5}
In this section, we shall derive the effective field theory describing the non-Hermitian Hubbard model in the honeycomb lattice,
where the non-Hermitian term explicitly breaks SU(2) spin-rotational symmetry. 
In the numerical simulation, the non-Hermitian term is introduced in the nearest-neighbor hopping term.
Linearizing the kinetic energy near the Dirac point, 
the low-energy effective theory is described a Dirac fermion field $\psi$ coupled with a complex scalar field $\phi=\phi_1+i\phi_2$,
\begin{equation}
\begin{aligned}
\mathcal{L} =&\mathcal{L}_{\psi} +\mathcal{L}_{\phi} +\mathcal{L}_{\psi\phi}, \\
\mathcal{L}_{\psi}=&
\bar{\psi} \big[\gamma_{0} \partial_{0}
+ v_F\gamma_{i} \partial_{i} +\delta \Gamma_i \partial_{i} \big] \psi 
,  \\
\mathcal{L}_{\phi}=& |\partial_{0}\phi|^2 +c^2|\partial_i\phi|^2 +m^2 |\phi|^2,  \\
\mathcal{L}_{\psi\phi}=& g \bar{\psi} (\phi_1 +i\gamma^5 \phi_2) \psi.  
\end{aligned}
\label{fullLagrangian}
\end{equation}
where the conjugate field is defined as $\bar{\psi}=\psi^{\dagger}\gamma_0$. 
The gamma matrices $\gamma_{\mu}$ satisfy Clifford algebra, $\{\gamma_{\mu},\gamma_{\nu}\}=2\delta_{\mu\nu}$.
A typical choice of the $\gamma$-matrices is given by $\gamma_0= s_3\otimes \sigma_3,\gamma_1= s_3\otimes \sigma_2,\gamma_2= -s_3\otimes \sigma_1$, $\gamma_3=s_1\otimes \sigma_0$,
where $s_{\mu}$ and $\sigma_{\mu}$ are Pauli matrices in the spin and sublattice space.
The chiral matrix $\gamma_5$ is defined as $\gamma_5=\gamma_0\gamma_1\gamma_2\gamma_3=-s_2\otimes \sigma_0$.
The Yukawa coupling can be written in a more symmetric form,
\begin{align}
\mathcal{L}_{\psi\phi}=& g \bar{\psi} (\phi_1 +i\gamma^5 \phi_2) \psi
= g \phi \bar{\psi} P_+ \psi 
+g \bar{\phi} \bar{\psi} P_- \psi 
\end{align}
with the project operators $P_+=\frac{1 +\gamma^5}{2},P_-=\frac{1 -\gamma^5}{2}$.
Since $\gamma_0=s_3\otimes \sigma_3$ and $i\gamma_0\gamma_5=-s_1\otimes \sigma_0$, such Yukawa coupling breaks the SU(2)-spin rotational symmetry down to the U(1) around the $s_2$ axis.
The non-Hermitian $\delta$-term only conserve this U(1) symmetry and the $\Gamma$ matrices are given by, $\Gamma_1=i\gamma_2 \gamma_5=s_1\otimes\sigma_1$, $\Gamma_2=-i\gamma_1 \gamma_5=s_1\otimes\sigma_2$.

Without the non-Hermitian term, the system belongs to the Gross-Neveu-Yukawa (GNY) XY university class, characterized by a Lorentz-invariant GNY chiral-XY quantum critical point (QCP) at finite coupling strength $g_c^2$.
The fermion velocity $v_F$ and boson velocity $c$ flow to the same value at this QCP.
In the following, we will introduce the non-Hermitian term $\delta$ as a small perturbation to this GNY-QCP and we can set $v_F=c=1$ for simplicity in the calculation.

We will study the one-loop renormalization group (RG) flow of the non-Hermitian $\delta$-term by momentum-shell RG method.
The bare propagators of the Dirac fermion and complex scalar field in the $D=d+1$ space-time dimension are given by
\begin{align}
\langle \psi(k) \bar{\psi}(k^{\prime}) \rangle 
=\frac{(2\pi)^{d+1}\delta(k-k^{\prime})}{i\gamma_{\mu} k_{\mu}},\quad
\langle \phi(k) \bar{\phi}(k^{\prime}) \rangle= \frac{(2\pi)^{d+1}\delta(k-k^{\prime})}{k^2}
\end{align}
The non-hermitian term and the Yukawa couping in Eq.(\ref{fullLagrangian}) are treated as perturbation,
\begin{align}
S_{\text{int}}=S_{\delta} +S_g
= +i\delta \int_k  \bar{\psi}_{k} \big(\Gamma_ik_i\big) \psi_k
+g \int_{k,q} \Big(\phi_q \bar{\psi}_{k+q} P_+ \psi_k
+ \bar{\phi}_q \bar{\psi}_{k} P_- \psi_{k+q} \Big)
\end{align}
The one-loop fermion self-energy correction is given by
\begin{align}
&\Sigma(k)
=\frac{g^2}{2}
\int_{q>} P_+ \langle \psi_{k-q} \bar{\psi}_{k-q}\rangle_> P_- \langle \phi_q\bar{\phi}_{q}\rangle_>
+\frac{g^2}{2} \int_{q}
P_- \langle \psi_{k+q} \bar{\psi}_{k+q}\rangle_> P_+ 
\langle \phi_q\bar{\phi}_{q} \rangle_>	\notag\\
=&-\frac{g^2}{2} \frac{2S_D}{D \Lambda^{4-D}} \frac{d\Lambda}{\Lambda}
\Big( P_+(ik_{\mu} \gamma_{\mu}) P_- +P_- (ik_{\mu} \gamma_{\mu}) P_+ \Big)
=-\frac{g^2}{2} \frac{2S_D}{D \Lambda^{4-D}} \frac{d\Lambda}{\Lambda} 
(ik_{\mu} \gamma_{\mu})
\end{align}
where $\Lambda$ is the momentum cutoff.
On the contrary, it is easy to check that the $\delta$-term will not contribute to the fermion field renormalization at one-loop level.
The effective fermion dispersion is given by $(ik_{\mu} \gamma_{\mu}) -\Sigma(k)$, and we only have the fermion field renormalization $Z_f$, with the anomalous dimension $\eta_f$ given by,
\begin{align}
Z_f= \sqrt{1+\eta_f},\quad
\eta_f \equiv \hat{g}^2= \frac{S_D}{D \Lambda^{4-D}} g^2,
\label{eq:Zf-etaf}
\end{align}
which is always positive semi-definite.

The one-loop correction to the $\delta$ term is given by,
\begin{align}
\delta S_{\delta}=& +i\frac{\delta g^2}{2} \int_{k}
\bar{\psi}_{k} P_+  \Big(\int_q \langle \psi_{k-q} \bar{\psi}_{k-q}\rangle  
\big(\Gamma_i(k-q)_i\big) 
\langle \psi_{k-q} \bar{\psi}_{k-q} \rangle \langle \phi_q \bar{\phi}_{q} \rangle \Big)
P_- \psi_{k}
	\notag\\
& +i\frac{\delta g^2}{2} \int_{k} 
\bar{\psi}_{k} P_- \Big( \int _q \langle \psi_{k+q} \bar{\psi}_{k+q} \rangle \big(\Gamma_i(k+q)_i\big) 
\langle \psi_{k+q} \bar{\psi}_{k+q} \rangle \langle \phi_q \bar{\phi}_{q} \rangle \Big)
P_+ \psi_k	\notag\\
=& +i\frac{\delta g^2}{2} \int_{k}
\bar{\psi}_{k} \big( P_+ I P_-+ P_-I P_+ \big)\psi_{k}
\end{align}
Here, the integral is
\begin{eqnarray}
\begin{aligned}
I_i=&\int_q \langle \psi_{k-q} \bar{\psi}_{k-q}\rangle  
\big(\Gamma_i(k-q)_i\big) 
\langle \psi_{k-q} \bar{\psi}_{k-q} \rangle \langle \phi_q \bar{\phi}_{q} \rangle 	\\
=&\int_q \frac{i\gamma_{\mu} q_{\mu}}{q_0^2 +\bm{q}^2} 
\Gamma_i q_i
\frac{i\gamma_{\nu} q_{\nu}}{q_0^2 +\bm{q}^2} \frac{1}{(k_0-q_0)^2 +(\bm{k}-\bm{q})^2}.
\end{aligned}
\end{eqnarray}
Expand the integral in order of $\bm{k}$, and we keep the linear order term,
\begin{eqnarray}
\begin{aligned}
I_1
=&+\int_q \frac{2q_{\mu} q_1 q_{\nu} q_{\rho}}{(q_0^2 +\bm{q}^2)^4} k_{\rho} (i\gamma_{\mu}) \Gamma_1 (i\gamma_{\nu}),	\\
I_2
=& +\int_q \frac{2q_{\mu} q_2 q_{\nu} q_{\rho}}{(q_0^2 +\bm{q}^2)^4} k_{\rho} (i\gamma_{\mu}) \Gamma_2 (i\gamma_{\nu}),
\end{aligned}
\end{eqnarray}
To be precise, we focus on the correction to the $k_1\Gamma$ ($\Gamma_1=(i\gamma_0 \gamma_1$) term, and a straightforward calculation gives,
\begin{align}
I_1(k_1) +I_2(k_1) 
=\frac{2S_D}{D\Lambda^{4-D}} \frac{d\Lambda}{\Lambda} (-\Gamma_1).
\end{align}
Here, the matrix part is calculated as,
\begin{align}
&P_+(\Gamma_1)P_- + P_-(\Gamma_1)P_+
=(\Gamma_1)P_-P_- + (\Gamma_1)P_+P_+
=\Gamma_1.
\end{align}
Together, the one-loop diagram to the $\delta$ term is 
\begin{align}
\delta S_{\delta}
=& -i\delta \hat{g}^2 dl \int_{k}
\bar{\psi}_{k} \Gamma_ik_i \psi_{k}.
\label{eq:delta}
\end{align}
Thus, the total one-loop renormalization correction to the $\delta$ comes from the fermion field renormalization in Eq.~(\ref{eq:Zf-etaf}) and one-loop diagram in Eq.~(\ref{eq:delta}),
\begin{align*}
\delta \rightarrow \big(\delta -\delta \hat{g}^2 dl\big)
\big(1 -\eta_f dl\big)
\end{align*}
Finally, the one-loop RG flow of the non-Hermitian parameter $\delta$ is given by,
\begin{align}
\frac{d\delta}{dl} =\big( -\hat{g}^2-\eta_f \big) \delta
=-2\hat{g}^2 \cdot\delta.
\end{align}
Here, the coupling strength $\hat{g}>0$ and takes finite value $\hat{g}_c^2$ at the GNY-XY QCP.
$\delta$ will flow to zero at this finite QCP.
We conclude that the non-Hermitian is an irrelevant perturbation at the Gross-Neveu-Yukawa-XY QCP, and the low-energy effective theory belongs to the same Hermitian Gross-Neveu-Yukawa-XY university class.

%\begin{thebibliography}{99}  
%\bibitem{Zerf-2017}Zerf, Nikolai, et al. "Four-loop critical exponents for the Gross-Neveu-Yukawa models." Physical Review D 96.9 (2017): 096010.  
%\end{thebibliography}

\end{widetext}
\end{document}